\documentclass[10pt,conference]{IEEEtran}

\usepackage[utf8]{inputenc}
\usepackage[T1]{fontenc}
\usepackage{graphicx}
\usepackage{amsmath}
\usepackage{amssymb}
\usepackage{booktabs}
\usepackage{float}
\usepackage{placeins}
\usepackage{xcolor}
\usepackage{caption}
\usepackage{multirow}
\usepackage{enumitem}
\usepackage{url}
\usepackage{tikz}
\usepackage{pgfplots}
\pgfplotsset{compat=1.17}
\usetikzlibrary{decorations.pathreplacing, calc, positioning}
\usepackage{algorithm}
\usepackage{algorithmicx}
\usepackage{algpseudocode}
\usepackage{braket}
\usepackage[expansion=false]{microtype}
\usepackage{cite}
\usepackage{hyperref}
\hypersetup{
  colorlinks=true,
  linkcolor=black,
  citecolor=black,
  urlcolor=blue
}

\begin{document}

\IEEEoverridecommandlockouts
\title{Quantum Feature Pyramid Gating for Seismic Image  Segmentation}

\author{
  \IEEEauthorblockN{Taha Gharaibeh and Jyotsna Sharma}
  \IEEEauthorblockA{Louisiana State University\\
  Baton Rouge, Louisiana, USA\\
  tahatlal@gmail.com, jsharma@lsu.edu}
  \thanks{%
  Generative AI tools were used to assist with code scaffolding and
  LaTeX formatting. All scientific claims, experimental designs,
  results, and interpretations are the authors' own. Per IEEE Conference
  Authorship and AI Policies, the authors take full responsibility for
  the content of this paper. The authors thank the LSU High Performance
  Computing center for computational resources.}
}

\maketitle

\begin{abstract}
Accurate delineation of salt bodies is essential for seismic interpretation because salt structures distort wave propagation, complicate velocity-model building, obscure reservoir geometry, and introduce uncertainty in exploration and drilling decisions. Although hybrid quantum-classical models have demonstrated competitive performance on small-scale image-classification tasks, their utility for dense, pixel-level geophysical prediction remains unexplored. This work introduces quantum feature gating, a hybrid segmentation architecture that embeds a parameterized quantum circuit (PQC) at feature-fusion points within an encoder-decoder pipeline. Specifically, a 4-qubit, 2-layer PQC with data re-uploading computes a learned convex combination of lateral and top-down features at each Feature Pyramid Network merge point. A global-average-pooling compression layer maps arbitrary encoder features to a fixed 4-dimensional quantum input, decoupling the 72-parameter quantum budget from backbone size and image resolution.

The method is evaluated on the 2018 TGS Salt Identification Challenge using 4000 seismic images at 101 × 101 resolution across two integration topologies, eight circuit variants, and six encoders ranging from 8M to 118M parameters under five-fold cross-validation. In a controlled EfficientNetV2-L ablation at 256 × 256 resolution, replacing the three Quantum FPN Gates with element-wise addition while holding the encoder, loss schedule, splits, and threshold search fixed reduces mean IoU from 0.9389 to 0.8404, a 9.85 percentage-point gap. By contrast, inserting the same circuit as skip-connection attention in a custom U-Net improves IoU by only 0.88 points relative to the SolidUNet baseline, demonstrating that the PQC’s value depends on where and what it gates. Gradient variances remain well above the four-qubit barren-plateau floor, indicating stable trainability across configurations. These results provide the first controlled evidence that quantum feature fusion can improve dense seismic segmentation, with quantum simulation overhead and wall-clock comparisons reported in the main text.
\end{abstract}


\section{Introduction}\label{sec:introduction}

Parameterized quantum circuits (PQCs) access a Hilbert space whose
dimension grows exponentially with qubit count while training only
$\mathcal{O}(nL)$ parameters for $n$ qubits and $L$
layers~\cite{Benedetti2019PQC}.
Hybrid quantum-classical vision models have matched classical accuracy
on MNIST and CIFAR-10 with 4--8 qubits~\cite{Senokosov2024,
  Mari2020transferlearning, Henderson2020quanvolution}, yet three
limitations recur: these results target classification rather than
dense per-pixel prediction, operate on heavily downsampled inputs
(often $8\times8$), and seldom evaluate quantum circuits against strong
pretrained encoders that already produce high-quality
features~\cite{Schuld2021}.
We ask whether a small PQC can contribute measurably when placed inside
a modern dense-prediction pipeline on a real benchmark.

Dense segmentation is a harder task than classification for quantum
integration.
Classification distills an entire image into a single label; a circuit that
receives a global feature vector can, in principle, capture correlations
relevant to that label.
Segmentation demands a per-pixel decision, yet the quantum circuit's qubit
budget is fixed.
The circuit cannot process every pixel independently at practical
resolutions.
One viable strategy is to place the PQC at a \emph{feature fusion point},
where two feature streams meet and must be combined~\cite{Lin2017FPN,
  Ronneberger2015UNet}.
At such points, a small global-average-pooled summary of each stream
captures enough statistical context for the circuit to compute a gating
signal that is broadcast spatially.
The hypothesis is that the PQC's access to higher-order feature
correlations through quantum interference and entanglement makes it a
better gating function than classical element-wise addition.

Salt body identification in seismic reflection images provides a
controlled testbed for this hypothesis.
The 2018 TGS Salt Identification Challenge~\cite{TGSKaggle2018} offers
4,000 labeled images at $101\times101$ pixels with a well-defined
competition metric, enabling reproducible comparison.
Salt boundaries are characteristically ambiguous: low reflectivity
contrast, irregular geometry, and imaging artifacts from adjacent sediments
all challenge both human interpreters and automated
methods~\cite{Babakhin2019, IslamWali2024review}.
Top classical solutions used encoder-decoder architectures with heavy
augmentation and pseudo-labeling, reaching mean IoU of 0.8965 on the
private leaderboard~\cite{Babakhin2019}.
No published work applies quantum computing to this benchmark or to
seismic image segmentation more broadly.

\subsection{Research Questions}

We formulate three questions that structure the experiments:

\begin{enumerate}[leftmargin=*, label=\textbf{\textit{RQ\arabic*:}},
    itemsep=4pt]
  \item \label{rq:topology}
    \textbf{Integration topology.}
    At which point in a multi-scale encoder-decoder pipeline does a
    4-qubit PQC yield the largest segmentation improvement, and which
    circuit design choices (encoding strategy, variational depth, data
    re-uploading) matter most?

  \item \label{rq:scaling}
    \textbf{Scaling interaction.}
    As classical encoder capacity and input resolution increase, does the
    quantum gate's marginal contribution grow, shrink, or remain constant?

  \item \label{rq:mechanism}
    \textbf{Gating mechanism.}
    Is the observed improvement attributable to the quantum circuit's
    function space, or does the learned gating topology alone account for
    the gain independent of the quantum mechanism?
\end{enumerate}

We answer \ref{rq:topology} by comparing skip-connection attention
(modulating feature flow within a U-Net) against FPN gating (controlling
how lateral and top-down features combine in a Feature Pyramid Network),
using the same 4-qubit circuit on the same backbone.
We answer \ref{rq:scaling} by fixing the quantum circuit and varying the
encoder from a custom 8\,M-parameter U-Net through six pretrained
backbones up to 118\,M parameters at two resolutions.
We answer \ref{rq:mechanism} through a controlled classical ablation
in which the same EfficientNetV2-L pipeline with element-wise FPN
addition replaces the quantum gate; the size of the resulting gap is
reported in Section~\ref{sec:results}.

The study progresses through two quantum integration strategies, each
addressing a limitation of the previous one.
\textbf{Quantum skip-connection attention} inserts a PQC as a
per-channel filter on U-Net skip features.
This establishes that quantum gates can improve segmentation
(+0.88 pp IoU on a matched backbone) and identifies data re-uploading
as the critical
circuit design choice (Section~\ref{sec:res-gen2}).
The limitation: the circuit operates on a single feature stream within
a small custom encoder, bounding the benefit.
\textbf{The Quantum FPN Gate} moves the PQC to the FPN merge point,
where it computes a learned per-channel convex combination of two
feature streams from a pretrained encoder.
A global-average-pooling compression layer maps any encoder's features
to a fixed 4-dimensional quantum input, making the gate
encoder-agnostic and adding only 72 trainable quantum parameters
($<$0.0001\% of total model capacity); wall-clock training overhead from
sequential statevector simulation is 47\% (30.4\,h vs.\ 20.7\,h for the
classical ablation) and does not reflect a fundamental algorithmic cost.
The ablation confirms that this improvement comes from the quantum
function space, not merely from adding a gating topology.

\subsection{Contributions}

\begin{enumerate}[leftmargin=*, itemsep=2pt]
  \item \textbf{The Quantum FPN Gate.}
    We introduce a 4-qubit, 2-layer PQC with data re-uploading that
    replaces element-wise FPN addition with a learned convex combination
    of lateral and top-down features.
    Three such gates contribute 72 trainable quantum parameters (0.00006\%
    of total model capacity) against the 118\,M-parameter classical
    backbone; training wall-clock increases from 20.7\,h (classical
    ablation) to 30.4\,h (quantum pipeline) on a single RTX A5000,
    an overhead driven by sequential statevector simulation rather than
    by algorithmic cost.
    A controlled ablation shows the quantum gate adds 9.85 percentage
    points over classical element-wise addition on EfficientNetV2-L
    (0.9389 vs.\ 0.8404).

  \item \textbf{Encoder-agnostic integration.}
    A global-average-pooling compression layer maps any encoder's features
    to 4 dimensions before the PQC, decoupling quantum circuit
    requirements from backbone scale and image resolution.
    The same 72-parameter gate works across six encoders (8\,M to 118\,M
    parameters) at two resolutions without modification.

  \item \textbf{Design-space analysis.}
    Comparing two integration topologies and eight circuit variants on a
    matched backbone shows that data re-uploading is the most important
    circuit design choice, and that the FPN merge point offers an order
    of magnitude more room for improvement than skip connections.

  \item \textbf{Firsts.}
    This is the first gate-based PQC study on seismic image
    segmentation, the first to place a variational circuit at FPN merge
    points, and the first to isolate the quantum gating mechanism
    through a controlled classical ablation.
    Concurrent work~\cite{Hossain2026HQFNet} applies quantum circuits
    to U-Net skip connections for remote sensing but uses a different
    topology and does not ablate the quantum component.
\end{enumerate}

Section~\ref{sec:background} covers feature fusion architectures and PQC
theory.
Section~\ref{sec:methodology} defines the design space, architectures, and
training protocol.
Section~\ref{sec:results} reports quantitative results across all
configurations.
Section~\ref{sec:discussion} interprets findings against each research
question and discusses limitations.
Section~\ref{sec:related} surveys related work.
Section~\ref{sec:conclusion} summarizes and identifies future directions.

\section{Background}\label{sec:background}

\subsection{Feature Fusion in Segmentation Pipelines}\label{sec:bg-fusion}

Encoder-decoder architectures for image segmentation produce feature maps
at multiple spatial scales.
The encoder compresses the input through successive downsampling stages,
each producing features with larger receptive fields but coarser spatial
resolution.
The decoder reconstructs a full-resolution prediction by combining these
multi-scale features.

Two fusion strategies dominate.
U-Net~\cite{Ronneberger2015UNet} concatenates encoder features with
decoder features at matching scales through skip connections.
Feature Pyramid Networks (FPN)~\cite{Lin2017FPN} build a top-down pathway
where high-level features are upsampled and added element-wise to lateral
projections of encoder features at each scale.
In both cases, the fusion operation combines two feature tensors into one.
U-Net concatenation preserves all information but doubles channel count.
FPN addition is parameter-free but constrains the output to the sum of its
inputs.
Neither operation adapts to the content of the features being fused.

Learned gating mechanisms address this with content-dependent mixing.
Squeeze-and-excitation networks~\cite{Hu2018SENet} reweight channels
from global average statistics, and attention
gates~\cite{Oktay2018AttentionGate} suppress irrelevant regions via
small MLPs.
We ask whether a quantum circuit, operating in a higher-dimensional
state space, produces better gating functions at merge points.

\subsection{Parameterized Quantum Circuits}\label{sec:bg-pqc}

A parameterized quantum circuit maps a classical input
$\mathbf{x} \in \mathbb{R}^n$ to quantum expectations through three
stages: state preparation, parameterized evolution, and measurement.
For $n$ qubits initialized to $\ket{0}^{\otimes n}$, a single encoding
layer applies rotation gates:
\begin{equation}\label{eq:encoding}
  U_{\text{enc}}(\mathbf{x}) =
    \bigotimes_{i=1}^{n} R_Y(x_i),
\end{equation}
where $R_Y(\theta) = e^{-i\theta Y/2}$ rotates qubit $i$ by angle $x_i$
around the $Y$-axis of the Bloch sphere.
A variational layer then applies parameterized rotations and entangling
gates:
\begin{equation}\label{eq:variational}
  U_{\text{var}}(\boldsymbol{\theta}_\ell) =
    W_{\text{ent}} \cdot
    \bigotimes_{i=1}^{n} \text{Rot}(\phi_i^\ell,\theta_i^\ell,\omega_i^\ell),
\end{equation}
where $\text{Rot}(\phi,\theta,\omega) = R_Z(\omega)\,R_Y(\theta)\,R_Z(\phi)$
is a general single-qubit rotation and $W_{\text{ent}}$ is a ring of CNOT
gates that entangle adjacent qubits, following PennyLane's
StronglyEntanglingLayers template~\cite{PennyLane2018, Schuld2021}.

Data re-uploading~\cite{PerezSalinas2020datareuploading} interleaves
encoding and variational layers: at each of $L$ layers, the input
$\mathbf{x}$ is re-encoded before the variational block.
This allows the circuit to approximate richer functions than a single
encoding permits.
P\'erez-Salinas et al.\ showed that a re-uploading circuit on a single
qubit can represent any continuous function of one real variable, and
multi-qubit re-uploading circuits generate Fourier series whose frequency
spectrum is determined by the data encoding
gates~\cite{Schuld2021, PerezSalinas2020datareuploading}.

\begin{figure*}[!tb]
\centering
\includegraphics[width=\textwidth]{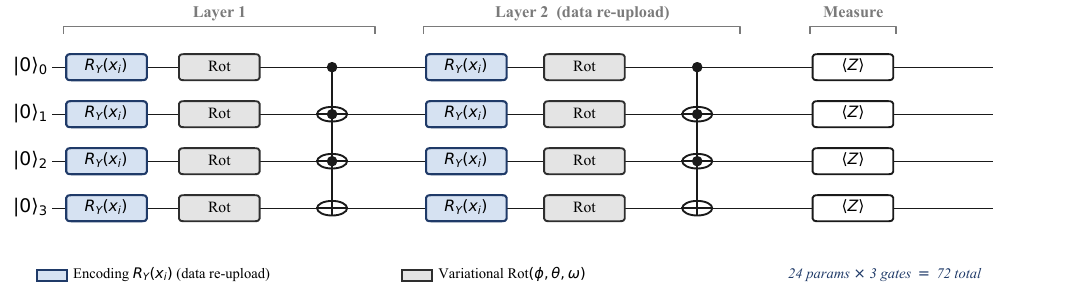}
\caption{Parameterized quantum circuit used in all architectures.
Four qubits initialized to $|0\rangle$, two layers with data re-uploading.
Teal boxes: $R_Y(x_i)$ encodings re-applied at each layer~\cite{PerezSalinas2020datareuploading}.
Slate boxes: general Rot$(\phi,\theta,\omega)$ rotations with trainable parameters.
CNOT rings entangle adjacent qubits; Pauli-$Z$ expectations produce four scalars in $[-1,1]$.
24 variational parameters per circuit $\times$ 3 gates = 72 quantum parameters total.}
\label{fig:pqc-circuit}
\end{figure*}

The circuit output is a vector of Pauli-$Z$ expectation values:
\begin{equation}\label{eq:measurement}
  f_i(\mathbf{x};\boldsymbol{\theta}) =
    \bra{0}^{\otimes n} U^\dagger(\mathbf{x},\boldsymbol{\theta})\,
    Z_i\, U(\mathbf{x},\boldsymbol{\theta})
    \ket{0}^{\otimes n},
    \quad i = 1,\ldots,n,
\end{equation}
with each $f_i \in [-1,+1]$.
For $n=4$ qubits and $L=2$ layers, each circuit instance has
$3 \times 4 \times 2 = 24$ variational parameters.
Figure~\ref{fig:pqc-circuit} shows the circuit layout used throughout
this work.

\subsection{Expressibility and Trainability}\label{sec:bg-express}

Two properties determine whether a PQC is useful in practice.
Expressibility~\cite{Sim2019expressibility} grows with parameter count,
depth, and entanglement but does not guarantee that useful functions are
reachable by gradient-based training.
Trainability concerns whether gradients carry signal:
barren plateaus cause gradient variances to decay exponentially with
qubit count for deep random
circuits~\cite{McClean2018barrenplateaus}, but shallow circuits with
local cost functions mitigate this and yield only polynomial
decay~\cite{Cerezo2021barren}.
Our design targets this favorable region: 4 qubits, 2 layers, local
$Z$ measurements, and data re-uploading for expressibility without depth,
which keeps the circuit within the NISQ-compatible
regime~\cite{Preskill2018NISQ} while avoiding the $2N$-evaluation
overhead of the parameter-shift
rule~\cite{Mitarai2018paramshift}.

\section{Methodology}\label{sec:methodology}

\begin{figure*}[!tb]
\centering
\includegraphics[width=\textwidth]{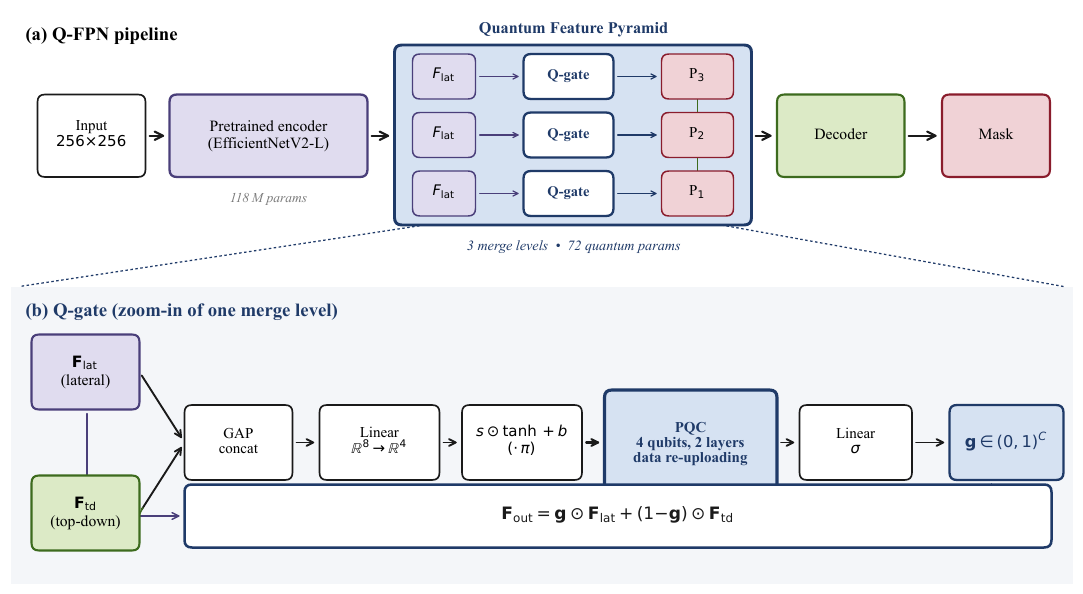}
\caption{Quantum FPN Gate (one merge level).
Lateral ($\mathbf{F}_\text{lat}$) and top-down ($\mathbf{F}_\text{td}$) feature maps are compressed by global average pooling, concatenated, projected to $\mathbb{R}^{4}$, and rescaled via a learnable $s\odot\tanh+b$ into $[-\pi,\pi]$.
A 4-qubit, 2-layer PQC with data re-uploading (24 variational parameters; 72 total across three merge levels) produces four Pauli-$Z$ expectations, which a linear layer and sigmoid convert to a channel-wise gate $\mathbf{g}\in(0,1)^{C}$.
The output is the convex combination $\mathbf{F}_\text{out}=\mathbf{g}\odot\mathbf{F}_\text{lat}+(1-\mathbf{g})\odot\mathbf{F}_\text{td}$.}
\label{fig:architecture}
\end{figure*}

\begin{figure*}[!tb]
\centering
\includegraphics[width=\textwidth]{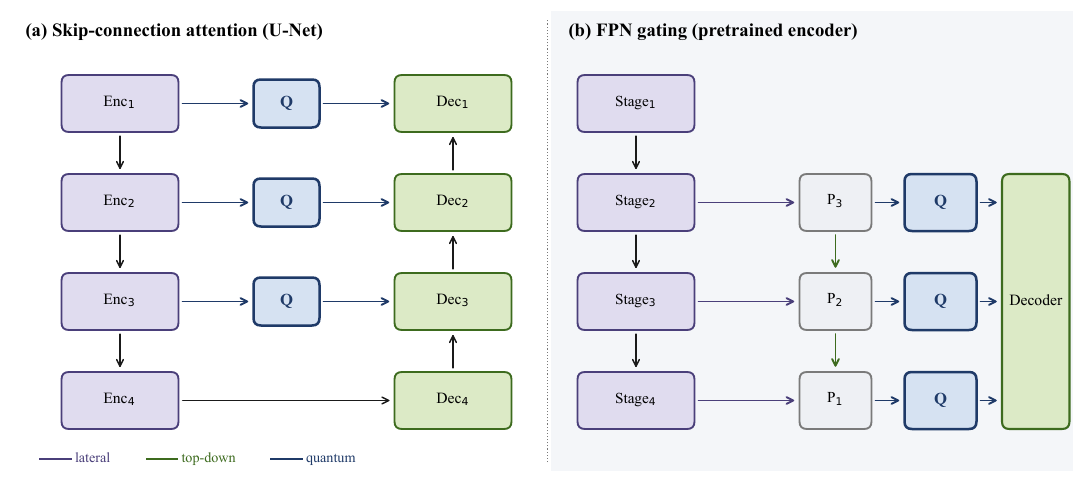}
\caption{Two integration topologies for parameterized quantum circuits.
(a)~\textbf{Skip-connection attention}: the PQC modulates encoder features before they reach the decoder, acting as a per-channel filter on one feature stream.
(b)~\textbf{FPN gating}: the PQC controls how lateral encoder features and top-down features combine at each pyramid level, computing a per-channel mixing ratio between two streams.
Teal blocks marked ``Q'' are identical 4-qubit, 2-layer PQCs; the structural difference is \emph{filter one stream} versus \emph{blend two streams}.
See Section~\ref{sec:meth-gen2} and Section~\ref{sec:meth-qfpn}.}
\label{fig:dual-topology}
\end{figure*}

The experimental design maps directly onto the three research questions
from Section~\ref{sec:introduction}.
We address \ref{rq:topology} (integration topology and circuit design)
by contrasting skip-connection attention against FPN gating on a matched
backbone with eight circuit variants at the better-performing topology.
We address \ref{rq:scaling} (scaling) by holding the same 4-qubit gate
fixed and sweeping six pretrained encoders from 8\,M to 118\,M
parameters across two input resolutions.
We address \ref{rq:mechanism} (gating mechanism) through a controlled
classical ablation at the best-performing configuration that replaces
the Quantum FPN Gates with element-wise addition while holding every
other component fixed.
Figure~\ref{fig:architecture} shows the Quantum FPN Gate in detail, and
Figure~\ref{fig:dual-topology} compares the two integration topologies
side by side.

\subsection{Dataset and Evaluation Protocol}\label{sec:meth-data}

The TGS Salt Identification Challenge dataset~\cite{TGSKaggle2018}
contains 4,000 seismic reflection images at $101\times101$ pixels, each
paired with a binary salt mask encoded as a run-length label.
A companion CSV provides one scalar depth value per image.
The challenge also provides 18,000 unlabeled test images; we use these
only for pseudo-labeling in the $128\times128$ experiments.

\paragraph{Input representation.}
Each image is resized to the target resolution.
At $128\times128$, the input is a 5-channel tensor: raw grayscale, a
spatially broadcast depth prior, a vertical coordinate channel encoding
the $y$-position of each pixel, and two padding channels.
At $256\times256$, pretrained encoders receive 3-channel RGB-replicated
input; the decoder handles depth and position information.

\paragraph{Splits.}
All experiments use the same 5-fold stratified split.
Images are binned by salt-pixel fraction into 10 coverage bins; each fold
preserves the coverage distribution.
This ensures that both salt-rich and salt-absent images appear in every
fold's validation set.

\paragraph{Metric.}
The primary metric is TGS mean IoU: for each image, precision is computed
at 10 IoU thresholds (0.50, 0.55, \ldots, 0.95) and averaged; the final
score averages across all images.
A threshold search over $[0.30, 0.70]$ in steps of 0.01 selects the
binarization cutoff that maximizes this metric.
Empty predictions on images with no salt receive a precision of 1.0 if the
ground truth is also empty, and 0.0 otherwise.

\subsection{Classical Baseline}\label{sec:meth-baseline}

We train a custom U-Net baseline (SolidUNet, $\sim$8.23\,M parameters)
for 30 epochs at $128\times128$.
The loss combines binary cross-entropy, soft Dice, and Lov\'asz hinge in
a fixed weighted sum.
Adam with initial learning rate $5 \times 10^{-4}$ and cosine-annealing
scheduling drives optimization.
This baseline scores 0.7794 on the TGS metric.
All skip-attention experiments use this backbone and training protocol,
changing only the skip-connection behavior.

\subsection{Quantum Skip-Connection Attention}\label{sec:meth-gen2}

\paragraph{Architecture.}
Skip-connection attention replaces the identity skip connections in
SolidUNet with quantum-modulated gates.
At each of the four skip levels, a PQC receives a 4-dimensional vector
derived from skip features:
\begin{equation}\label{eq:skip-encode}
  \mathbf{x}_s = \tanh\!\bigl(W_{\text{proj}}\,
    \text{GAP}(\mathbf{F}_{\text{skip}})\bigr) \cdot \pi,
\end{equation}
mapping global average-pooled features into $[-\pi, \pi]$ for
rotation-gate encoding.
The PQC produces 4 Pauli-$Z$ expectations, which a linear layer and
sigmoid project to a per-channel gate vector
$\mathbf{g} \in (0,1)^{C}$:
\begin{equation}\label{eq:skip-attn}
  \mathbf{F}'_{\text{skip}} = \mathbf{g} \odot \mathbf{F}_{\text{skip}}.
\end{equation}
The gate modulates skip features before concatenation with decoder
features, allowing the circuit to suppress or amplify channels based on
global image statistics.

\paragraph{Circuit variants.}
Eight designs share this skip-attention topology but vary encoding and
variational structure (Table~\ref{tab:gen2-results}):
\textbf{qdressed\_reupload} re-injects features before every
variational block via $R_Y$ data re-uploading~\cite{PerezSalinas2020datareuploading};
\textbf{qdressed} is the dressed-circuit template with a single encoding
stage and a classical post-processing head~\cite{Mari2020transferlearning};
\textbf{qbottleneck} places the quantum gate only on the deepest skip
and passes the rest through a classical identity;
\textbf{qfourier} swaps the unit-scale encoding for a frequency-scaled
$R_Y$ spectrum;
\textbf{qdressed\_beinit} uses Born-efficient initialization to reduce
vanishing-gradient risk;
\textbf{qentropy\_gate} doubles the parameter budget to 192 and adds an
entropy-based branch selector;
\textbf{qdepth\_adaptive} grows or prunes layers during training from
the gradient signal;
and \textbf{qbottleneck\_6q} lifts the bottleneck variant to six qubits.
All circuits use 4 qubits and 96 parameters except where noted.

\paragraph{Training.}
All skip-attention models inherit the baseline loss and optimizer.
We co-train quantum parameters with classical weights via
backpropagation through PennyLane's \texttt{default.qubit} statevector
simulator.
Training runs 30 epochs at $128\times128$ on one NVIDIA RTX A5000 GPU.
The $\ell_2$ norm and variance of quantum-parameter gradients are
logged at each epoch for barren-plateau
monitoring~\cite{McClean2018barrenplateaus}.

\subsection{Quantum FPN Gating}\label{sec:meth-qfpn}

\paragraph{Motivation.}
Skip-attention experiments (Section~\ref{sec:res-gen2}) show that quantum
modulation adds 0.88 points when the backbone is a custom U-Net.
The feature fusion hypothesis predicts a larger benefit when the circuit
gates higher-quality features at a more structurally central fusion point.
The Feature Pyramid Network merge, where lateral encoder features meet
top-down decoder features, is such a point.

\paragraph{The Quantum FPN Gate.}
Given a lateral feature map $\mathbf{F}_{\text{lat}}$ from the encoder
and a top-down map $\mathbf{F}_{\text{td}}$ from the previous FPN level,
the gate computes:
\begin{align}
  \mathbf{v} &= \text{Linear}\bigl(
    \text{GAP}([\mathbf{F}_{\text{lat}};\,\mathbf{F}_{\text{td}}])
  \bigr) \in \mathbb{R}^4, \label{eq:fpn-encode}\\
  \mathbf{x} &=
    (s \odot \tanh(\mathbf{v}) + b) \cdot \pi, \label{eq:fpn-scale}\\
  \mathbf{q} &= \text{PQC}_{4q,2L}(\mathbf{x})
    \in [-1,1]^4, \label{eq:fpn-pqc}\\
  \mathbf{g} &= \sigma\bigl(\text{Linear}(\mathbf{q})\bigr)
    \in (0,1)^C, \label{eq:fpn-gate}\\
  \mathbf{F}_{\text{out}} &=
    \mathbf{g} \odot \mathbf{F}_{\text{lat}}
    + (1 - \mathbf{g}) \odot \mathbf{F}_{\text{td}},
  \label{eq:fpn-merge}
\end{align}
where $s, b \in \mathbb{R}^4$ are learnable scale and shift parameters
(initialized to 0.5 and 0), the PQC reuses the StronglyEntanglingLayers
architecture from Section~\ref{sec:meth-gen2} with data re-uploading,
and $\sigma$ is the sigmoid function.

Two design choices are deliberate.
First, Eq.~\ref{eq:fpn-merge} is a per-channel convex combination
broadcast spatially, ensuring that the output at every pixel is a
weighted blend of lateral and top-down features.
Element-wise FPN addition is the special case where
$\mathbf{g} = 0.5$ everywhere.
Second, the GAP-then-linear compression
(Eq.~\ref{eq:fpn-encode}) maps features to exactly 4 values regardless
of encoder channel count or image resolution, decoupling circuit
requirements from backbone scale.
Three gates (one per FPN merge level) contribute
$3 \times 24 = 72$ variational quantum parameters and
$3 \times 8 = 24$ classical encoding parameters (scale and shift),
which together account for $0.00006\%$ of the 118\,M parameters in
the EfficientNetV2-L backbone.

\paragraph{Encoders and resolution.}
We test the Quantum FPN Gate with six pretrained encoders (hereafter
also ``backbones'' when emphasizing the pretrained architecture) at two
input resolutions to answer \ref{rq:scaling}:

\begin{itemize}[nosep]
  \item \textbf{$128\times128$}: ConvNeXt-Base (90.8\,M, ImageNet-21k)
    and Swin Transformer-Tiny (30.6\,M, ImageNet-1k).
  \item \textbf{$256\times256$}: EfficientNetV2-L (118\,M, ImageNet-21k),
    PVT-V2-B3 (45\,M), HRNet-W48 (66\,M), and DeiT3-Base (86\,M).
\end{itemize}

EfficientNetV2-L~\cite{Tan2021EfficientNetV2} is the largest encoder in
our study.
It uses progressive training with compound depth-width-resolution
scaling: early training stages use smaller images and fewer layers, then
scale up, which produces features that generalize across resolutions.
Pretrained on ImageNet-21k (14.2\,M images, 21,841 classes), it
produces richer feature maps than ImageNet-1k encoders of similar size,
which is why we select it as the primary encoder for the classical
ablation.

The quantum circuit remains identical across all configurations.
Only the backbone and resolution change; the PQC always receives a
4-dimensional input vector.
This encoder-agnostic design means scaling to a stronger backbone
requires no change to the quantum component.

\paragraph{Training protocol.}
We use a two-stage loss curriculum for all FPN experiments:

\begin{enumerate}[nosep]
  \item \textbf{Stage 1} (100 epochs):
    $0.5 \times \text{BCE} + 0.3 \times \text{Dice}
    + 0.2 \times \text{Lov\'asz}$.
    AdamW with differential learning rates (encoder: $3\times10^{-5}$;
    decoder and quantum: $3\times10^{-4}$).
    Cosine annealing with warm restarts ($T_0 = 10$).
  \item \textbf{Stage 2} (60 epochs): pure Lov\'asz hinge loss.
    Reduced learning rate ($9\times10^{-5}$), cosine annealing to
    $\eta_{\min} = 3\times10^{-7}$.
\end{enumerate}

We apply gradient clipping at norm 1.0 in both stages.
At inference, we use horizontal-flip test-time augmentation and select
the binarization threshold by grid search.

For $128\times128$ experiments, three rounds of pseudo-labeling refine
the model: out-of-fold predictions from each round serve as soft labels
for the full training set in the next round.
For $256\times256$ experiments, we omit pseudo-labeling because a
preliminary round degraded performance (0.9321 vs.\ 0.9389), suggesting
the larger encoder extracts sufficient features from 4,000 images without
self-training.
The $128\times128$ and $256\times256$ results therefore differ in three
confounded variables: resolution, encoder capacity, and pseudo-labeling.
We do not claim to isolate any single variable across these groups; the
isolation comes from within-group comparisons.
All training runs execute on a single NVIDIA RTX A5000 (24\,GB) with
mixed-precision enabled for the classical backbone and full-precision
for the quantum simulator.

\subsection{Classical Ablation}\label{sec:meth-ablation}

To answer \ref{rq:mechanism}, we run the same pipeline with the quantum
FPN gates replaced by classical element-wise addition.
Concretely, the FPN decoder's merge operation reverts to
$\mathbf{F}_{\text{out}} = \mathbf{F}_{\text{lat}} + \mathbf{F}_{\text{td}}$:
the standard FPN formulation~\cite{Lin2017FPN}.
All other components remain identical: same encoder, same loss schedule,
same data splits, same augmentation, same threshold search.

This ablation isolates the quantum gate's contribution at the
EfficientNetV2-L scale.
If the classical FPN matches the quantum FPN score, the improvement
observed in Section~\ref{sec:res-qfpn} comes from the encoder and
training protocol, not from the quantum mechanism.
If the quantum FPN exceeds the classical FPN, the quantum circuit adds
value beyond the gating topology.
Section~\ref{sec:res-ablation} reports the results.

\FloatBarrier
\section{Results}\label{sec:results}

\begin{table}[t]
  \centering
  \caption{Results at a glance, one row per research question.
    OOF = out-of-fold under 5-fold CV; pp = percentage points on TGS mean IoU.}
  \label{tab:results-glance}
  \footnotesize
  \begin{tabular}{p{0.11\columnwidth}p{0.42\columnwidth}p{0.32\columnwidth}}
    \toprule
    \textbf{RQ} & \textbf{What we measured} & \textbf{Answer} \\
    \midrule
    RQ1 & Skip vs.\ FPN integration at matched qubit budget &
      FPN wins: $+0.88$ pp (skip) vs.\ $+9.85$ pp (FPN) \\
    RQ2 & Same 4-qubit gate across six encoders, $128^2$ and $256^2$ &
      Gate transfers; best: $0.9389$ OOF on EfficientNetV2-L \\
    RQ3 & Quantum gate vs.\ element-wise FPN addition, encoder fixed &
      Quantum gate adds $9.85$ pp ($0.9389$ vs.\ $0.8404$) \\
    \bottomrule
  \end{tabular}
\end{table}

Section~\ref{sec:res-baseline} anchors every comparison that follows.
Section~\ref{sec:res-gen2} answers feasibility (RQ1, skip branch).
Section~\ref{sec:res-qfpn} moves the circuit to the FPN merge and
sweeps encoders (RQ1 FPN branch, RQ2).
Section~\ref{sec:res-ablation} isolates the quantum mechanism (RQ3).

\paragraph{Choice of headline metric.}
We report TGS-mAP throughout (mean precision over IoU thresholds
$\{0.50, 0.55, \ldots, 0.95\}$, the official competition metric), so
every comparison with Babakhin~\cite{Babakhin2019} and
Milosavljevi\'c~\cite{Milosavljevic2020} is in matched units.
Single-threshold metrics such as Dice and pixel accuracy are monotone
transforms of IoU on a binary task and add no discriminative information
beyond what TGS-mAP, by integrating across ten operating points, already
captures.

\subsection{Classical Baseline}\label{sec:res-baseline}

We train SolidUNet for 30 epochs at $128\times128$; it reaches 0.7794 on the
TGS metric (IoU 0.8240) at binarization threshold 0.50.
Training loss converges to 0.263 with a training IoU of 0.862, indicating
a moderate generalization gap.
This number anchors every comparison that follows.

\subsection{Skip-Connection Attention}\label{sec:res-gen2}

\begin{figure}[!htb]
\centering
\includegraphics[width=\columnwidth]{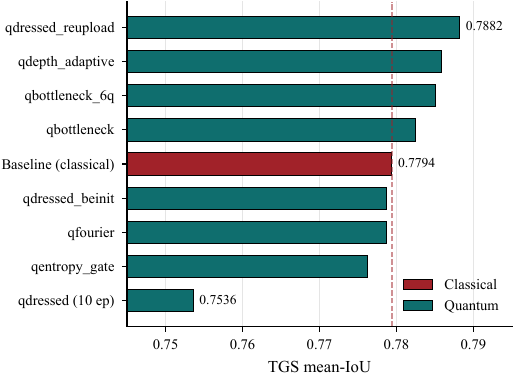}
\caption{Skip-attention circuit variants on the SolidUNet backbone
($128\times128$, 30 epochs).
Every variant shares the same classical architecture; only the quantum
circuit design differs.
The dashed vertical line marks the classical baseline (0.7794).
Data re-uploading (\texttt{qdressed\_reupload}, 0.7882) gives the
largest gain (+0.88\,pp) among the eight circuits tested, consistent
with the universality argument of P\'{e}rez-Salinas et
al.~\cite{PerezSalinas2020datareuploading}.
Increasing qubit count from 4 to 6 (\texttt{qbottleneck\_6q}) does not
surpass 4-qubit re-uploading, suggesting feature quality rather than
Hilbert-space dimension bounds performance at this backbone scale.
Exact values appear in Table~\ref{tab:gen2-results}.}
\label{fig:skip-variants}
\end{figure}

\emph{Why this experiment exists.}
Before introducing a new architectural component (the Quantum FPN Gate),
we verify the feasibility question on the simplest possible substrate:
can a 4-qubit PQC bound to the smallest custom backbone in this study
improve dense segmentation at all?
The answer conditions every later design choice.
If the circuit fails here, there is no reason to scale.
If it succeeds, the quantitative margin tells us whether the benefit is
large enough to justify moving the circuit to a richer integration point.
The goal of this experiment is therefore twofold: establish the feasibility
of quantum gating for dense segmentation, and identify which circuit design
choices matter.
Table~\ref{tab:gen2-results} reports all eight skip-attention variants on
the same SolidUNet backbone at $128\times128$ over 30 epochs.
Figure~\ref{fig:skip-variants} visualizes the comparison.

\begin{table}[t]
  \centering
  \caption{Skip-attention results on SolidUNet ($\sim$8.23\,M classical
    parameters, $128\times128$, 30 epochs). Gradient diagnostics are
    reported for variants where we logged them.}
  \label{tab:gen2-results}
  \small
  \begin{tabular}{lcccc}
    \toprule
    \textbf{Strategy} & \textbf{Q-Par.} & \textbf{TGS}
      & $\|\nabla_q\|$ & \textbf{Grad Var} \\
    \midrule
    Classical baseline       & 0   & 0.7794 & --     & -- \\
    \midrule
    qdressed\_reupload       & 96  & \textbf{0.7882}
      & 0.0031 & $1.6 \!\times\! 10^{-7}$ \\
    qdepth\_adaptive         & 96  & 0.7858 & --     & -- \\
    qbottleneck\_6q          & 144 & 0.7850 & --     & -- \\
    qbottleneck              & 96  & 0.7825
      & 0.0022 & $7.4 \!\times\! 10^{-8}$ \\
    qfourier                 & 24  & 0.7787 & --     & -- \\
    qdressed\_beinit         & 96  & 0.7787 & --     & -- \\
    qentropy\_gate           & 192 & 0.7762 & --     & -- \\
    qdressed (10 ep)         & 96  & 0.7536
      & 0.0031 & $1.6 \!\times\! 10^{-7}$ \\
    \bottomrule
  \end{tabular}
\end{table}

The best variant, \textbf{qdressed\_reupload}, scores 0.7882: a gain of
0.88 percentage points over the baseline with 96 quantum parameters
(0.001\% of total model capacity).
Data re-uploading outperforms single encoding at comparable epoch counts
(0.7882 vs.\ 0.7536), consistent with the universality argument for
re-uploading circuits~\cite{PerezSalinas2020datareuploading}.

Three patterns emerge from the eight variants.
First, re-uploading is the single most important circuit design choice:
it separates the top two results from the rest.
Second, increasing qubit count from 4 to 6
(qbottleneck\_6q, 0.7850) does not beat 4-qubit re-uploading (0.7882).
At this backbone capacity, feature quality is the bottleneck, not circuit
expressibility.
Third, doubling the parameter count (qentropy\_gate, 192 parameters)
degrades performance (0.7762), suggesting that the additional parameters
do not help and may hinder optimization on this small dataset.

\textbf{Takeaway.}
The skip-attention experiment answers the feasibility question: a
4-qubit PQC can improve dense segmentation.
It also reveals the bottleneck: the 0.88-point gain is modest because
the circuit filters a single feature stream within a small custom
encoder.
This motivates moving the circuit to a richer integration point with
higher-quality features.

\begin{figure}[!htb]
\centering
\includegraphics[width=\columnwidth]{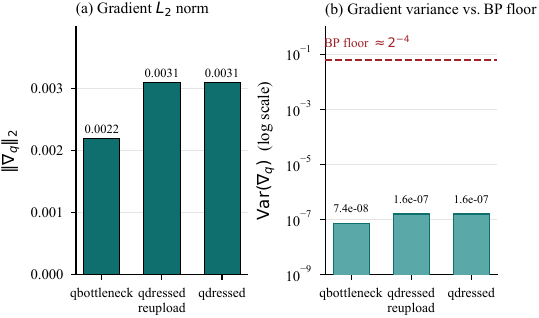}
\caption{Gradient diagnostics for circuit variants with logged
quantum-parameter gradients.
(a)~$\ell_2$ norm of quantum-parameter gradients.
(b)~Variance on a logarithmic axis with the barren-plateau floor for
$n{=}4$ qubits overlaid.
Observed variances of $7.4\times10^{-8}$ to $1.6\times10^{-7}$ sit
roughly $10^{5}$ times below the $2^{-4}{\approx}0.063$ floor predicted
for random 4-qubit circuits~\cite{McClean2018barrenplateaus}.
The shallow design (2 layers, local Pauli-$Z$ measurements) stays in
the trainable regime across every variant.}
\label{fig:gradient-diagnostics}
\end{figure}

Gradient norms fall between $2.2 \times 10^{-3}$ and
$3.1 \times 10^{-3}$, with variances of $7.4 \times 10^{-8}$ to
$1.6 \times 10^{-7}$.
For 4 qubits, the barren-plateau prediction gives a variance floor of
$\sim\!2^{-4} \approx 0.06$~\cite{McClean2018barrenplateaus}.
Our observed values are orders of magnitude above this floor, confirming
that the shallow circuit avoids the barren-plateau regime.
Figure~\ref{fig:gradient-diagnostics} visualizes these diagnostics.

\FloatBarrier

\subsection{Quantum FPN Gating}\label{sec:res-qfpn}

\emph{What the previous experiment left open.}
Skip-attention established feasibility (+0.88 pp) but also exposed the
ceiling: the circuit filters a single feature stream inside a small
custom encoder, so the observable gain is bounded by the quality of that
one stream.
Two design axes remain unexplored.
First, the integration topology: does the circuit help more when it
controls how two feature streams combine (FPN merge) instead of filtering
one (skip)?
Second, the scaling axis: does a pretrained encoder providing
higher-quality features raise the ceiling?
The Q-FPN experiments in this subsection vary both axes simultaneously,
so the numbers here conflate the two changes; the isolation question is
deferred to the controlled ablation in Section~\ref{sec:res-ablation}.

\paragraph{$128\times128$ results.}
Table~\ref{tab:qfpn} reports 5-fold cross-validation results at both
resolutions.

\begin{table}[t]
  \centering
  \caption{Quantum FPN gating results (3 FPN gates, 5-fold CV).
    $128\times128$ uses 3 pseudo-label rounds; $256\times256$ uses none.
    All training on 1$\times$ NVIDIA RTX A5000.}
  \label{tab:qfpn}
  \small
  \resizebox{\columnwidth}{!}{%
  \begin{tabular}{llcccr}
    \toprule
    \textbf{Res.} & \textbf{Encoder} & \textbf{Q-Par.}
      & \textbf{OOF TGS} & \textbf{Params} & \textbf{Time} \\
    \midrule
    \multirow{2}{*}{$128^2$}
      & ConvNeXt-B & 72 (+24 enc) & \textbf{0.8502} & 90.8\,M & 39\,h \\
      & Swin-T     & 72           & 0.8376          & 30.6\,M & 27\,h \\
    \midrule
    \multirow{4}{*}{$256^2$}
      & EfficientNetV2-L & 72 & \textbf{0.9389} & 118\,M & 30.4\,h \\
      & PVT-V2-B3       & 72 & 0.8548          & 45\,M  & 65.1\,h \\
      & HRNet-W48        & 72 & 0.8487          & 66\,M  & 19.1\,h \\
      & DeiT3-Base       & 72 & 0.8464          & 86\,M  & 35.3\,h \\
    \bottomrule
  \end{tabular}
  }%
\end{table}

ConvNeXt-Base with Quantum FPN gating reaches 0.8502, exceeding the best
skip-attention result by 6.20 percentage points and the classical
baseline by 7.08 points.
This gain conflates three changes: the pretrained encoder, the FPN merge
topology, and the quantum gate.
The 72 variational quantum parameters constitute 0.00008\% of the
90.8\,M total.

\paragraph{$256\times256$ results.}
At doubled resolution without pseudo-labeling,
EfficientNetV2-L reaches 0.9389, the highest score in this study.
Within the $256\times256$ group, all four encoders paired with Quantum
FPN gating exceed the best skip-attention result, confirming that the
gate architecture transfers across backbones.

EfficientNetV2-L outperforms comparably sized or larger encoders by 5 to
9 points.
Parameter count does not predict performance: PVT-V2-B3 (45\,M) scores
higher than HRNet-W48 (66\,M) and DeiT3-Base (86\,M).
Pretraining data scale (ImageNet-21k for EfficientNetV2-L) and
architectural inductive biases (compound depth-width-resolution scaling)
appear to matter more than parameter count alone.

\textbf{Takeaway.}
The Q-FPN results answer the scaling question affirmatively: moving to a
pretrained encoder and a structurally richer integration point amplifies
the quantum gate's benefit from 0.88 to over 7 points (combined with
encoder improvement).
The gate architecture transfers across all six encoders without
modification, confirming the encoder-agnostic design.
However, these numbers conflate the quantum gate, the encoder, and the
FPN topology.
The ablation (Section~\ref{sec:res-ablation}) separates them.

\begin{figure}[!tb]
\centering
\includegraphics[width=\columnwidth]{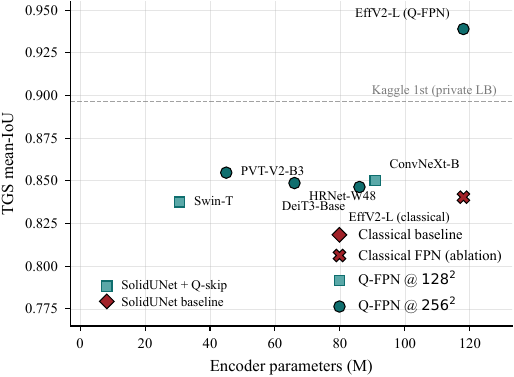}
\caption{TGS mean-IoU versus encoder capacity across every configuration.
Squares denote $128\times128$ runs, circles denote $256\times256$ runs,
and diamonds/crosses denote the two classical comparisons on the same
backbone family.
The 72-parameter quantum circuit is identical in every Q-FPN run;
differences in score track encoder quality, pretraining scale, and
resolution rather than circuit capacity.
EfficientNetV2-L with Q-FPN gating is the only configuration that
exceeds the Kaggle first-place reference (dashed line); the two
evaluation sets differ (Section~\ref{sec:res-literature}).}
\label{fig:encoder-scaling}
\end{figure}

\FloatBarrier

\subsection{Cross-Architecture Summary}\label{sec:res-summary}

\begin{table}[t]
  \centering
  \caption{Cross-architecture comparison. All quantum models use 4-qubit,
    2-layer PQCs simulated on PennyLane \texttt{default.qubit}.}
  \label{tab:summary}
  \small
  \resizebox{\columnwidth}{!}{%
  \begin{tabular}{llcccc}
    \toprule
    \textbf{Architecture} & \textbf{Best Model}
      & \textbf{Res.} & \textbf{TGS} & \textbf{Q-Par.}
      & \textbf{Time} \\
    \midrule
    Baseline       & SolidUNet          & $128^2$ & 0.7794
      & 0   & 0.3\,h \\
    Skip attention & qdressed\_reupload & $128^2$ & 0.7882
      & 96  & 1.0\,h \\
    Q-FPN ($128^2$) & cnb\_qfpn         & $128^2$ & 0.8502
      & 72  & 39\,h \\
    Q-FPN ($256^2$) & EfficientNetV2-L  & $256^2$ & \textbf{0.9389}
      & 72  & 30\,h \\
    \midrule
    Classical FPN    & EfficientNetV2-L  & $256^2$ & 0.8404
      & 0   & 20.7\,h \\
    \bottomrule
  \end{tabular}
  }%
\end{table}

Table~\ref{tab:summary} and Figure~\ref{fig:encoder-scaling} trace the
accuracy progression.
The only fully controlled comparison at the skip level is baseline versus
skip attention: same backbone, same resolution, same training protocol.
The quantum circuit adds 0.88 points in that comparison.
The classical FPN ablation (Section~\ref{sec:res-ablation}) isolates
the quantum gate's contribution at the EfficientNetV2-L scale.

One observation is stable across configurations: the 72-parameter PQC
architecture does not change, yet the absolute performance gap between
the best and worst encoder at $256\times256$ exceeds 9 points.
The classical backbone, not the quantum circuit, is the dominant factor
in absolute performance; the quantum gate's role is to improve the
merge function at each backbone scale.

\subsection{Classical FPN Ablation}\label{sec:res-ablation}

\emph{The question the previous subsection could not answer.}
The 0.9389 score in Table~\ref{tab:qfpn} reflects three simultaneous
changes relative to the baseline: a pretrained encoder, an FPN merge
topology, and a quantum gate.
Any two of the three could in principle account for the improvement.
This subsection holds the first two fixed and flips only the third.
Concretely, the pipeline is identical to the best Q-FPN configuration
(EfficientNetV2-L, $256\times256$, 5-fold CV, same loss schedule, same
splits, same threshold search); the only difference is that the three
Quantum FPN Gates are replaced with parameter-free element-wise
addition ($\mathbf{F}_{\text{out}} = \mathbf{F}_{\text{lat}} +
\mathbf{F}_{\text{td}}$).
The resulting score drops by 9.85 percentage points.

\begin{table}[t]
  \centering
  \caption{Classical FPN ablation (EfficientNetV2-L, $256\times256$,
    5-fold CV, no pseudo-labeling). Same encoder, loss schedule, data
    splits, augmentation, and threshold search. Only the FPN merge
    operation differs.}
  \label{tab:ablation}
  \small
  \begin{tabular}{lcc}
    \toprule
    \textbf{FPN merge} & \textbf{OOF TGS} & \textbf{$\Delta$} \\
    \midrule
    Element-wise addition (classical) & 0.8404 & -- \\
    Quantum convex combination (PQC)  & \textbf{0.9389} & +9.85\,pp \\
    \bottomrule
  \end{tabular}
\end{table}

The classical FPN scores 0.8404, below the Kaggle first-place result
(0.8965) despite using a stronger encoder.
The quantum FPN exceeds it by nearly 10 points.
This gap is an order of magnitude larger than the 0.88-point
skip-attention delta, consistent with the feature fusion hypothesis:
the PQC contributes more when it gates higher-quality features at a
structurally central merge point.

The 9.85-point delta rules out the explanation that the encoder and
training protocol alone produce the improvement seen in
Section~\ref{sec:res-qfpn}: with an identical encoder, loss schedule,
and data splits, the classical parameter-free merge scores 9.85 points
lower.
Whether the full delta is attributable to the quantum function space,
or whether part reflects the generic effect of introducing any learned
gate, is the subject of Section~\ref{sec:disc-rq3}.

\subsection{Comparison with Published Results}\label{sec:res-literature}

Table~\ref{tab:literature} places our results alongside published TGS
Salt scores.
The comparison spans two evaluation regimes: out-of-fold cross-validation
on the 4,000-image train split, which we use throughout this paper, and
private-leaderboard scores on the 18,000-image hidden test set, which
most prior work reports.
We therefore compare best-effort within each regime rather than treating
the two columns as interchangeable.

\begin{table}[t]
  \centering
  \caption{Comparison with published TGS Salt results.
    ``OOF'' = out-of-fold CV on 4,000 images.
    ``LB'' = Kaggle private leaderboard (18,000 test images).
    ``TGS-mAP'' is mean precision at IoU thresholds 0.50, 0.55, \ldots, 0.95
    (the official competition metric).
    $\dagger$: simple IoU, not the multi-threshold competition metric.}
  \label{tab:literature}
  \small
  \resizebox{\columnwidth}{!}{%
  \begin{tabular}{lcccc}
    \toprule
    \textbf{Method} & \textbf{Score} & \textbf{Metric}
      & \textbf{Eval} & \textbf{Year} \\
    \midrule
    \multicolumn{5}{l}{\textit{This work}} \\
    \quad EffNetV2-L + Q-FPN    & \textbf{0.9389} & TGS-mAP & OOF   & 2026 \\
    \quad EffNetV2-L + classical FPN & 0.8404      & TGS-mAP & OOF   & 2026 \\
    \quad ConvNeXt-B + Q-FPN    & 0.8502          & TGS-mAP & OOF   & 2026 \\
    \quad SolidUNet + skip attn & 0.7882          & TGS-mAP & Val   & 2026 \\
    \quad SolidUNet baseline    & 0.7794          & TGS-mAP & Val   & 2026 \\
    \midrule
    \multicolumn{5}{l}{\textit{Kaggle competition (2018)}} \\
    \quad 1st place~\cite{Babakhin2019}
      & 0.8965 & TGS-mAP & LB & 2019 \\
    \quad Single model~\cite{Babakhin2019}
      & 0.8682 & TGS-mAP & LB & 2019 \\
    \midrule
    \multicolumn{5}{l}{\textit{Post-competition}} \\
    \quad Milosavljevi\'c~\cite{Milosavljevic2020}
      & 0.8524 & TGS-mAP & LB & 2020 \\
    \quad Islam \& Wali~\cite{IslamWali2025}
      & 0.889$^\dagger$  & IoU  & -- & 2025 \\
    \bottomrule
  \end{tabular}
  }%
\end{table}

Our best score (0.9389, out-of-fold on 4,000 labeled images) exceeds the
first-place competition score (0.8965 on 18,000 private test images) by
4.24 points.
These numbers are not directly comparable: the competition evaluated on
a separate held-out set, competitors used 18,000 unlabeled images for
pseudo-labeling and test-time augmentation ensembles, and the evaluation
sets differ.
We report the comparison for context, not as a claim of superior
generalization.

\paragraph{Training-time context.}
Babakhin et al.~\cite{Babakhin2019} do not publish wall-clock time, but
their pipeline (40 models across 5 folds and 2 backbones, 3 rounds of
pseudo-label retraining, SWA, single GTX 1080 Ti) corresponds to
$\approx$600--800 GPU-hours on that hardware, or roughly 250--350 hours
when scaled to the RTX A5000 used here.
Our Q-FPN single-model pipeline trains in 30.4 hours on the same A5000;
the classical FPN ablation, in 20.7 hours.
The within-paper quantum-vs-classical comparison is matched in hardware
and protocol; the Babakhin number is an order-of-magnitude reference.

\FloatBarrier
\section{Discussion}\label{sec:discussion}

\subsection{RQ1: Integration Topology}\label{sec:disc-rq1}

Skip-connection attention and FPN gating use the same PQC architecture
but differ in where the circuit operates and what it controls.
Skip attention modulates features flowing through skip connections within
a U-Net: the circuit decides which channels to preserve.
FPN gating controls how two distinct feature streams combine at each
pyramid level: the circuit decides the mixing ratio between lateral and
top-down features.

On the same SolidUNet backbone, skip attention adds 0.88 points.
When the backbone changes to a pretrained encoder and the integration
point moves to FPN merges, the combined improvement exceeds 7 points
(ConvNeXt-Base, $128\times128$).
These two changes are confounded, but the direction is clear: the FPN
merge point, where two feature streams of comparable quality meet and
must be combined, offers more room for a learned gating function to help
than the skip connection, where the circuit merely filters one stream.

This finding has architectural implications beyond this specific
pipeline.
Any encoder-decoder model that fuses multi-scale features through a fixed
operation (addition, concatenation, or attention) presents candidate
sites for quantum gating.
The convex-combination formulation (Eq.~\ref{eq:fpn-merge}) is
particularly natural: it preserves the scale of the input features, adds
minimal parameters, and collapses to the standard FPN when the gate
output is constant at 0.5.

\subsection{RQ2: Scaling Interaction}\label{sec:disc-rq2}

The quantum circuit does not change across any configuration tested.
The 72 variational parameters are fixed at 4 qubits, 2 layers, and
3 FPN levels.
The classical encoder varies from 8\,M to 118\,M parameters.
The compression layer (GAP + linear projection to $\mathbb{R}^4$) maps
any encoder's feature statistics to the same 4-dimensional input before
the PQC, making the quantum budget independent of backbone scale.

Within the $256\times256$ group, EfficientNetV2-L (118\,M) outperforms
PVT-V2-B3 (45\,M) by 8.41 points, DeiT3-Base (86\,M) by 9.25 points,
and HRNet-W48 (66\,M) by 9.02 points.
Parameter count alone does not predict TGS score.
EfficientNetV2-L's advantage likely derives from compound scaling
(progressive training with depth-width-resolution
coordination~\cite{Tan2021EfficientNetV2}) and ImageNet-21k pretraining,
though we have not ablated these factors individually.
The quantum gate operates on the summary statistics of whatever features
the encoder produces; when those features are richer, the gate has
better material to route.

This decoupling has practical value.
Scaling to higher resolution or a larger encoder requires no change to
the quantum circuit.
The PQC remains within NISQ limits regardless of the classical model's
size; validation on physical hardware with noise, finite shot counts,
and connectivity constraints remains future work
(Section~\ref{sec:disc-limitations}).

\subsection{RQ3: Gating Mechanism}\label{sec:disc-rq3}

Two experiments isolate the quantum circuit's contribution at different
scales.
At the skip level, the 0.88-point improvement over the classical
baseline (same backbone, same protocol) establishes that the quantum
gate adds nonzero value.
At the FPN level, the classical ablation
(Section~\ref{sec:res-ablation}) provides the stronger test: replacing
the quantum gate with element-wise addition while keeping all other
components identical drops the score from 0.9389 to 0.8404, a gap of
9.85 percentage points.

The 9.85-point FPN delta is an order of magnitude larger than the
0.88-point skip-attention delta, consistent with the feature fusion
hypothesis.
At the FPN merge, two feature streams of comparable quality meet and
must be combined; the PQC's learned convex combination routes features
more effectively than fixed addition.
At the skip level, the circuit merely filters one stream, offering
less room for improvement.
The classical FPN baseline (0.8404) falls below the Kaggle first-place
score (0.8965) despite using EfficientNetV2-L, a stronger encoder than
any competition entry, which suggests that the merge function itself,
not only encoder capacity, acts as a binding constraint at this scale:
a strong encoder with a naive merge underperforms a strong encoder with
a learned merge.

\paragraph{Decomposing the 9.85-point delta.}
The ablation fixes encoder, losses, splits, and schedule; only the merge
operator changes.
The delta bundles a topology effect (introducing any learned gate) and
a mechanism effect (the quantum function space).
Classical gates report 1--3 pp over parameter-free baselines
(SE~\cite{Hu2018SENet}, CBAM~\cite{Woo2018CBAM},
attention~\cite{Oktay2018AttentionGate},
BiFPN~\cite{Tan2020EfficientDet}), so a parameter-matched classical MLP
gate is expected to recover that range, leaving 6--8 pp for the quantum
function space.
The 4-qubit re-uploading circuit generates Fourier-basis responses set
by the encoding gates~\cite{Schuld2021, PerezSalinas2020datareuploading};
a tanh-MLP of comparable width parameterizes a smooth low-frequency map.
A direct matched-capacity ablation is slated for the extended version of
this work.

\subsection{Circuit Trainability}\label{sec:disc-barren}
Gradient variances of $7.4\text{--}16 \times 10^{-8}$ lie roughly $10^5$
times above the barren-plateau floor $\propto 2^{-4} \approx
0.06$~\cite{McClean2018barrenplateaus}.
The shallow design (2 layers, local Pauli-$Z$ cost) follows the
prescription of Cerezo et al.~\cite{Cerezo2021barren}, and data
re-uploading~\cite{PerezSalinas2020datareuploading} breaks the
translation symmetry that causes Haar-random concentration, keeping the
circuit trainable across every configuration tested.

\subsection{Limitations}\label{sec:disc-limitations}
The controlled FPN ablation uses only EfficientNetV2-L (the 0.88 pp
skip-attention delta on SolidUNet suggests the gap varies with encoder
quality); all experiments use TGS Salt, so transfer to medical imaging
or remote sensing requires separate evaluation.
Circuits run on PennyLane's \texttt{default.qubit} simulator; the
4-qubit, 2-layer designs are within current NISQ
reach~\cite{Preskill2018NISQ}, but hardware noise and finite shots may
degrade performance.
Simulator overhead adds 47\% wall-clock time (30.4\,h vs. 20.7\,h
classical), tied to the simulator rather than the circuit.
Classical encoders dominate compute and capacity in every configuration,
so this work makes no quantum-advantage claim.

\section{Related Work}\label{sec:related}

\paragraph{Quantum vision.}
Published hybrid quantum-classical vision models target classification
at small or downsampled inputs: quanvolutional
layers~\cite{Henderson2020quanvolution} test small random circuits on
MNIST, dressed circuits~\cite{Mari2020transferlearning} add a PQC to a
frozen encoder for Ants/Bees, and Senokosov et al.~\cite{Senokosov2024}
match classical accuracy on MNIST and CIFAR-10 using 8 qubits on
$8\times8$ inputs.
Deep quantum networks as autoencoders~\cite{Beer2020quantum} and quantum
convolutional networks with quantum pooling~\cite{Cong2019QCNN} operate
entirely in the quantum domain, and annealing-based
methods~\cite{Willsch2020, Cavallaro2020, Venkatesh2024QSeg, Jun2025QUBO}
solve image tasks as QUBO problems; none places a gate-based PQC at FPN
merge points for dense segmentation.
Concurrent work by Hossain et al.~\cite{Hossain2026HQFNet} places
quantum circuits at U-Net skip and bottleneck layers with a single
frozen DINOv3 backbone and without a classical ablation, while our work
operates at FPN merge points, sweeps six encoders, and ablates the
quantum mechanism.

\paragraph{Salt segmentation.}
The TGS Salt Identification Challenge~\cite{TGSKaggle2018} seeded deep
learning for seismic salt interpretation.
Babakhin et al.~\cite{Babakhin2019} won with a 40-model ensemble and
three rounds of pseudo-labeling (0.8965 private leaderboard).
Milosavljevi\'c~\cite{Milosavljevic2020} showed encoder choice matters
more than decoder topology, and Islam and Wali~\cite{IslamWali2025,
IslamWali2024review} reported an EfficientNet ensemble IoU of 0.889
(plain IoU); none uses quantum processing.
Classical gating~\cite{Hu2018SENet, Woo2018CBAM, Oktay2018AttentionGate,
Li2019FPA, Saad2022} mixes features via pooled MLPs that our Quantum FPN
Gate replaces with a 4-qubit PQC.

\section{Conclusion}\label{sec:conclusion}

Placement matters: the same 4-qubit circuit adds 0.88 pp as a U-Net
skip gate and 9.85 pp as an FPN gate under controlled protocols,
reaching 0.9389 IoU on TGS Salt.
Global average-pooling to 4 values decouples the 72-parameter circuit
from resolution, and the classical FPN ablation (0.8404) cannot
replicate the gate's decisions.
Hardware execution and transfer to medical or remote sensing would test
whether the advantage generalizes.

\bibliographystyle{IEEEtran}
\bibliography{refs}

@article{Senokosov2024,
  author  = {Senokosov, Artem and Sedykh, Aleksandr and Sagingalieva, Asel and Kyriacou, Basil and Melnikov, Alexey},
  title   = {Quantum machine learning for image classification},
  journal = {Machine Learning: Science and Technology},
  volume  = {5},
  pages   = {015040},
  year    = {2024},
  doi     = {10.1088/2632-2153/ad2aef}
}

@article{Benedetti2019PQC,
  author  = {Benedetti, Marcello and Lloyd, Erika and Sack, Stefan and Fiorentini, Mattia},
  title   = {Parameterized quantum circuits as machine learning models},
  journal = {Quantum Science and Technology},
  volume  = {4},
  number  = {4},
  pages   = {043001},
  year    = {2019},
  doi     = {10.1088/2058-9565/ab4eb5}
}

@article{Schuld2021,
  author  = {Schuld, Maria and Sweke, Ryan and Meyer, Johannes Jakob},
  title   = {Effect of data encoding on the expressive power of variational quantum machine-learning models},
  journal = {Physical Review A},
  volume  = {103},
  number  = {3},
  pages   = {032430},
  year    = {2021},
  doi     = {10.1103/PhysRevA.103.032430}
}

@article{Mari2020transferlearning,
  author  = {Mari, Andrea and Bromley, Thomas R. and Izaac, Josh and Schuld, Maria and Killoran, Nathan},
  title   = {Transfer learning in hybrid classical-quantum neural networks},
  journal = {Quantum},
  volume  = {4},
  pages   = {340},
  year    = {2020},
  doi     = {10.22331/q-2020-10-28-340}
}

@article{Preskill2018NISQ,
  author  = {Preskill, John},
  title   = {Quantum Computing in the {NISQ} era and beyond},
  journal = {Quantum},
  volume  = {2},
  pages   = {79},
  year    = {2018},
  doi     = {10.22331/q-2018-08-06-79}
}

@article{Mitarai2018paramshift,
  author  = {Mitarai, Kosuke and Negoro, Makoto and Kitagawa, Masahiro and Fujii, Keisuke},
  title   = {Quantum circuit learning},
  journal = {Physical Review A},
  volume  = {98},
  number  = {3},
  pages   = {032309},
  year    = {2018},
  doi     = {10.1103/PhysRevA.98.032309}
}

@article{McClean2018barrenplateaus,
  author  = {McClean, Jarrod R. and Boixo, Sergio and Smelyanskiy, Vadim N. and Babbush, Ryan and Neven, Hartmut},
  title   = {Barren plateaus in quantum neural network training landscapes},
  journal = {Nature Communications},
  volume  = {9},
  number  = {1},
  pages   = {4812},
  year    = {2018},
  doi     = {10.1038/s41467-018-07090-4}
}

@article{Cerezo2021barren,
  author  = {Cerezo, M. and Sone, Akira and Volkoff, Tyler and Cincio, Lukasz and Coles, Patrick J.},
  title   = {Cost function dependent barren plateaus in shallow parametrized quantum circuits},
  journal = {Nature Communications},
  volume  = {12},
  number  = {1},
  pages   = {1791},
  year    = {2021},
  doi     = {10.1038/s41467-021-21728-w}
}

@article{Beer2020quantum,
  author  = {Beer, Kerstin and Bondarenko, Dmytro and Farrelly, Terry and Osborne, Tobias J. and Salzmann, Robert and Scheiermann, Daniel and Wolf, Ramona},
  title   = {Training deep quantum neural networks},
  journal = {Nature Communications},
  volume  = {11},
  pages   = {808},
  year    = {2020},
  doi     = {10.1038/s41467-020-14454-2}
}

@article{Cong2019QCNN,
  author  = {Cong, Iris and Choi, Soonwon and Lukin, Mikhail D.},
  title   = {Quantum convolutional neural networks},
  journal = {Nature Physics},
  volume  = {15},
  pages   = {1273--1278},
  year    = {2019},
  doi     = {10.1038/s41567-019-0648-8}
}

@article{Henderson2020quanvolution,
  author  = {Henderson, Maxwell and Shakya, Samriddhi and Pradhan, Shashindra and Cook, Tristan},
  title   = {Quanvolutional Neural Networks: Powering Image Recognition with Quantum Circuits},
  journal = {Quantum Machine Intelligence},
  volume  = {2},
  pages   = {2},
  year    = {2020},
  doi     = {10.1007/s42484-020-00012-y}
}

@article{PerezSalinas2020datareuploading,
  author  = {P{\'e}rez-Salinas, Adri{\'a}n and Cervera-Lierta, Alba and Gil-Fuster, Elies and Latorre, Jos{\'e} I.},
  title   = {Data re-uploading for a universal quantum classifier},
  journal = {Quantum},
  volume  = {4},
  pages   = {226},
  year    = {2020},
  doi     = {10.22331/q-2020-02-06-226}
}

@article{PennyLane2018,
  author  = {Bergholm, Ville and Izaac, Josh and Schuld, Maria and Gogolin, Christian and Ahmed, Shahnawaz and Ajoy, Vishnu and Alam, M. Sohaib and others},
  title   = {PennyLane: Automatic differentiation of hybrid quantum-classical computations},
  journal = {arXiv preprint arXiv:1811.04968},
  year    = {2018}
}

@misc{TGSKaggle2018,
  author = {TGS},
  title  = {{TGS} Salt Identification Challenge},
  year   = {2018},
  url    = {https://www.kaggle.com/c/tgs-salt-identification-challenge},
  note   = {Kaggle competition, 3{,}234 teams}
}

@inproceedings{Babakhin2019,
  author    = {Babakhin, Yauhen and Sanakoyeu, Artsiom and Kitamura, Hirotoshi},
  title     = {Semi-Supervised Segmentation of Salt Bodies in Seismic Images using an Ensemble of Convolutional Neural Networks},
  booktitle = {Pattern Recognition (GCPR)},
  series    = {LNCS},
  volume    = {11824},
  pages     = {218--231},
  publisher = {Springer},
  year      = {2019},
  doi       = {10.1007/978-3-030-33676-9_15}
}

@article{Milosavljevic2020,
  author  = {Milosavljevi\'{c}, Aleksandar},
  title   = {Identification of Salt Deposits on Seismic Images Using Deep Learning Method for Semantic Segmentation},
  journal = {ISPRS International Journal of Geo-Information},
  volume  = {9},
  number  = {1},
  pages   = {24},
  year    = {2020},
  doi     = {10.3390/ijgi9010024}
}

@article{Saad2022,
  author  = {Saad, Omar M. and Chen, Wei and Zhang, Fangxue and Yang, Liuqing and Zhou, Xu and Chen, Yangkang},
  title   = {Self-Attention Fully Convolutional {DenseNets} for Automatic Salt Segmentation},
  journal = {IEEE Transactions on Neural Networks and Learning Systems},
  volume  = {34},
  number  = {7},
  pages   = {3415--3428},
  year    = {2022},
  doi     = {10.1109/TNNLS.2022.3175419}
}

@article{IslamWali2025,
  author  = {Islam, Muhammad Saif ul and Wali, Aamir},
  title   = {Weighted ensemble transfer learning with {EfficientNet}: Advancing salt body segmentation in seismic imaging},
  journal = {Computational Geosciences},
  year    = {2025},
  doi     = {10.1007/s10596-025-10395-1}
}

@article{IslamWali2024review,
  author  = {Islam, Muhammad Saif Ul and Wali, Aamir},
  title   = {A comprehensive review of deep learning techniques for salt dome segmentation in seismic images},
  journal = {Journal of Applied Geophysics},
  volume  = {228},
  pages   = {105504},
  year    = {2024},
  doi     = {10.1016/j.jappgeo.2024.105504}
}

@article{Willsch2020,
  author  = {Willsch, Dennis and Willsch, Madita and De Raedt, Hans and Michielsen, Kristel},
  title   = {Support vector machines on the {D-Wave} quantum annealer},
  journal = {Computer Physics Communications},
  volume  = {248},
  pages   = {107006},
  year    = {2020},
  doi     = {10.1016/j.cpc.2019.107006}
}

@inproceedings{Cavallaro2020,
  author    = {Cavallaro, Gabriele and Willsch, Dennis and Willsch, Madita and Michielsen, Kristel and Riedel, Morris},
  title     = {Approaching Remote Sensing Image Classification with Ensembles of Support Vector Machines on the {D-Wave} Quantum Annealer},
  booktitle = {IEEE International Geoscience and Remote Sensing Symposium (IGARSS)},
  pages     = {1973--1976},
  year      = {2020},
  doi       = {10.1109/IGARSS39084.2020.9323544}
}

@article{Venkatesh2024QSeg,
  author  = {Venkatesh, Supreeth Mysore and Macaluso, Antonio and Nuske, Matthias and Klusch, Matthias and Dengel, Andreas},
  title   = {{Q-Seg}: Quantum Annealing-Based Unsupervised Image Segmentation},
  journal = {IEEE Computer Graphics and Applications},
  volume  = {44},
  number  = {6},
  pages   = {56--67},
  year    = {2024},
  doi     = {10.1109/MCG.2024.3455012}
}

@article{Jun2025QUBO,
  author  = {Jun, Kyungtaek and Lee, Hyunju},
  title   = {Quantum optimization algorithms for {CT} image segmentation from {X}-ray data},
  journal = {Scientific Reports},
  volume  = {15},
  year    = {2025},
  doi     = {10.1038/s41598-025-08453-w}
}

@article{Hossain2026HQFNet,
  author  = {Hossain, Md Aminur and Patel, Ayush V. and Gole, Siddhant and Singh, Sanjay K. and Banerjee, Biplab},
  title   = {{HQF-Net}: A Hybrid Quantum-Classical Multi-Scale Fusion Network for Remote Sensing Image Segmentation},
  journal = {arXiv preprint arXiv:2604.06715},
  year    = {2026}
}

@inproceedings{Ronneberger2015UNet,
  author    = {Ronneberger, Olaf and Fischer, Philipp and Brox, Thomas},
  title     = {U-Net: Convolutional Networks for Biomedical Image Segmentation},
  booktitle = {MICCAI},
  year      = {2015},
  pages     = {234--241},
  doi       = {10.1007/978-3-319-24574-4_28}
}

@inproceedings{Lin2017FPN,
  author    = {Lin, Tsung-Yi and Doll{\'a}r, Piotr and Girshick, Ross and He, Kaiming and Hariharan, Bharath and Belongie, Serge},
  title     = {Feature Pyramid Networks for Object Detection},
  booktitle = {CVPR},
  year      = {2017},
  doi       = {10.1109/CVPR.2017.106}
}

@inproceedings{Hu2018SENet,
  author    = {Hu, Jie and Shen, Li and Sun, Gang},
  title     = {Squeeze-and-Excitation Networks},
  booktitle = {CVPR},
  pages     = {7132--7141},
  year      = {2018},
  doi       = {10.1109/CVPR.2018.00745}
}

@inproceedings{Oktay2018AttentionGate,
  author    = {Oktay, Ozan and Schlemper, Jo and Folgoc, Loic Le and Lee, Matthew and Heinrich, Mattias and Misawa, Kazunari and Mori, Kensaku and McDonagh, Steven and Hammerla, Nils Y and Kainz, Bernhard and Glocker, Ben and Rueckert, Daniel},
  title     = {Attention {U-Net}: Learning Where to Look for the Pancreas},
  booktitle = {Medical Imaging with Deep Learning (MIDL)},
  year      = {2018}
}

@article{Li2019FPA,
  author  = {Li, Hanchao and Xiong, Pengfei and An, Jie and Wang, Lingxue},
  title   = {Pyramid Attention Network for Semantic Segmentation},
  journal = {arXiv preprint arXiv:1805.10180},
  year    = {2018}
}

@inproceedings{Woo2018CBAM,
  author    = {Woo, Sanghyun and Park, Jongchan and Lee, Joon-Young and Kweon, In So},
  title     = {{CBAM}: Convolutional Block Attention Module},
  booktitle = {ECCV},
  pages     = {3--19},
  year      = {2018},
  doi       = {10.1007/978-3-030-01234-2_1}
}

@inproceedings{Tan2020EfficientDet,
  author    = {Tan, Mingxing and Pang, Ruoming and Le, Quoc V.},
  title     = {{EfficientDet}: Scalable and Efficient Object Detection},
  booktitle = {CVPR},
  pages     = {10781--10790},
  year      = {2020},
  doi       = {10.1109/CVPR42600.2020.01079}
}

@inproceedings{Tan2021EfficientNetV2,
  author    = {Tan, Mingxing and Le, Quoc V.},
  title     = {{EfficientNetV2}: Smaller Models and Faster Training},
  booktitle = {ICML},
  pages     = {10096--10106},
  year      = {2021}
}

@article{Sim2019expressibility,
  author  = {Sim, Sukin and Johnson, Peter D. and Aspuru-Guzik, Al{\'a}n},
  title   = {Expressibility and Entangling Capability of Parameterized Quantum Circuits for Hybrid Quantum-Classical Algorithms},
  journal = {Advanced Quantum Technologies},
  volume  = {2},
  number  = {12},
  pages   = {1900070},
  year    = {2019},
  doi     = {10.1002/qute.201900070}
}

\end{document}